\documentstyle[12pt,epsf,aaspp4]{article}
\begin{document}

\title{DETERMINATION OF THE HUBBLE CONSTANT USING A TWO-PARAMETER
LUMINOSITY CORRECTION FOR TYPE Ia SUPERNOVAE}
\author{Robert Tripp}
\affil{Institute for Nuclear and Particle Astrophysics, Physics Division\\
Lawrence Berkeley National Laboratory, Berkeley, CA 94720}
\and
\author{David Branch}
\affil{Department of Physics and Astronomy, University of Oklahoma, Norman,
OK 73019}

\clearpage

\begin{abstract}
In this paper, we make a comprehensive determination of the Hubble
constant $H_0$ by using two parameters---the $B-V$ color and the
rate of decline $\Delta m_{15}$---to simultaneously standardize
the luminosities of all nearby Cepheid-calibrated type Ia supernovae
(SNe Ia) and those of a larger, more distant sample of 29 SNe Ia.  Each
group is treated in as similar a manner as possible in order to avoid
systematic effects.  A simultaneous $\chi ^2$ minimization yields
a standardized absolute luminosity of the Cepheid-calibrated
supernovae as well as the Hubble constant obtained from the more
distant sample.  We find $H_0 = 62$ km s$^{-1}$ Mpc$^{-1}$ and
a standardized absolute magnitude of -19.46.
The sensitivity of $H_0$ to a metallicity dependence
of the Cepheid-determined distances is investigated.  The total
uncertainty $\delta H_0$, dominated by uncertainties in the primary
Cepheid distance indicator, is estimated to be 5 km s$^{-1}$
Mpc$^{-1}$.
\end{abstract}


\clearpage

\section{INTRODUCTION}

One requirement for measuring the Hubble constant using SNe Ia
as standard candles is a sample of well-measured distant
supernovae.  They should be distant enough that their measured
redshifts are dominated by the Hubble flow, but not so distant
that the still-uncertain dynamics associated with deceleration
due to the mass density of the universe and possible acceleration
due to conjectured repulsive forces are important.  A broad range
of redshift $z$ from 0.01 to 0.2 is suitable, with $z$ about
0.05 being optimum.  For this purpose the carefully measured and
uniformly analyzed Cal\'an-Tololo collection of 29 SNe Ia  (Hamuy et
al. 1996)  covering the range $0.01 < z < 0.1$ is very well-suited.
The collection has a  spread in relative luminosity covering
more than 1.5 magnitudes, but it has been shown that a two-parameter
luminosity correction using the color and decline rate of each
supernova is both necessary and sufficient to standardize them to
a common luminosity (Tripp 1998).

A second requirement for measuring $H_0$ is to establish the absolute
luminosity of SNe Ia that have been standardized in the above
way.  At the present time there are seven nearby (typically an order
of magnitude closer) galaxies that have hosted SNe~Ia whose galaxy
distances have been determined using Cepheid variables measured by
the Hubble Space Telescope.  This includes the recently discovered
SN 1998bu in the Leo I group galaxy NGC 3368 whose Cepheid distance
had already been measured.  The Cepheid-determined distances each
have a typical accuracy of about 5\%, plus an additional overall
uncertainty of about 7\% associated with the absolute distance
scale for Cepheid variables.

These seven galaxies have hosted eight recorded SNe Ia.  The oldest,
SN 1895B, was observed photographically in only one color so cannot
be used here.  Three of the others preceded the use of CCDs in
astronomy and, despite being well-measured by the techniques of
the day, are not of the quality that can now be achieved.  Thus we also
consider an expanded list containing three more recent SNe Ia which,
although not found in Cepheid-calibrated galaxies, are very near on
the sky to such a galaxy or are generally considered to be within
groups of galaxies that have at least one member that has been
so calibrated.  Results are presented both with and without these
additional supernovae.

An extensive bibliograpy on the use of Type Ia supernovae for measuring the
Hubble constant
can be found in the recent review by Branch (1998).

\section{PROCEDURE}

We treat the 29 distant SNe of Hamuy et al. (1996) following the method of
Tripp (1998), except that we
allow the absolute magnitude to vary in a simultaneous fit with the
Cepheid-calibrated supernovae.  We
calculate the Hubble constant $H_0$ for each distant supernova using
\begin{equation}
\log H_0 = { {M_B - B + 52.38} \over 5 } + \log {{1 - q_0 + q_0 z - (1-q_0)
\sqrt{ 1 + 2 q_0 z } } \over { q_0^2}}
\end{equation}
Here $B$ and $M_B$ are the apparent and absolute blue magnitudes at maximum
light, while $q_0$ is the
deceleration parameter and $z$ is the measured red shift.  For the 29
distant SNe, $B$ and $z$ are tabulated
by Hamuy et al. (1996).  These supernovae are not so remote $(z \le 0.1)$
that the still uncertain value of
$q_0$ leads to a significant uncertainty.  Conforming to the current
evidence of Riess et al. (1998) and
Perlmutter et al. (1998) for a negative $q_0$, we fix it at -0.45 (see
Section 4).As shown previously (Tripp
1998), in order to fit the data of Hamuy et al. (1996) within their quoted
errors, the above absolute
magnitude $M_B$ for each supernova must be adjusted according to both its
rate of decline $\Delta
m_{15}$
and its color $(B - V)$.  Here $B$ and $V$ are the maximum apparent blue
and visual magnitudes.
Empirically, linear dependences prove  to be more than adequate for both
corrections.  We therefore write
for $M_B$ in equation (1)
\begin{equation}
M_B = <M_B^0> + b ( \Delta m_{15} - 1.05) + R (B - V)
\end{equation}
\noindent
The parameter $R$ incorporates both intrinsic color differences between the
SNe and any reddening caused
by dust in the host galaxies.  These two effects are often difficult to
distinguish, particularly for the more
distant supernovae.  The reddening parameter due to dust alone, usually
denoted by $R_B$, should be about
4 if the dust surrounding extragalactic supernovae is similar to dust in
the Milky Way.\footnote{
However, recent measurements of extinction from diffuse Galactic cirrus
clouds (Szomoru \& Guhathakurta 1999) find $R_B$ values $\lesssim 3$ due
perhaps to a
smaller average grain size.  Since SNe Ia are not closely associated with
star formation it may well be that cirrus values are relevant for them.  In
any event, all SNe with evidence of strong dust obscuration are removed
from the sample for the case where we apply a color
selection to the data.}

The weighted average value $<M_B^0>$ in equation (2) is obtained from the
Cepheid-calibrated SNe in
the following manner.  We use a similar dependence to standardize each of
them to the value it would
have if
$\Delta m_{15} = 1.05$ and $B - V = 0$ for that supernova.  Thus the
corrected absolute magnitude
becomes
\begin{equation}
M_B^0 = M_B - b (\Delta m_{15} - 1.05) - R (B - V)
\end{equation}

A least-squares fit of the Cepheid-calibrated SNe gives the weighted
average $<M_B^0>$ appearing in
equation (2), along with the $\chi ^2$ for the fit.  Both are functions of
$b$ and $R$.  The value 1.05 in
equations (2) and (3) is the average
$\Delta m_{15}$ for 13 SNe Ia compiled by Branch et al. (1996).  Chosen for
convenience, the choice is
arbitrary and has no effect on $H_0$, although it will directly affect
$<M_B^0>$.  The same remarks apply
to the arbitrary choice of a
$<B-V> = 0 $ subtraction in the color term.

For each Cal\'an-Tololo distant supernova, we use equation. (1)  to
evaluate $H_0$ using the value of
$M_B$ from
equation (2) with $b$ and $R$ as parameters.  The uncertainty in $H_0$ for
each supernova is obtained by
combining in quadrature the quoted errors $\delta B$, $\delta m_{15}$, and
$\delta (B - V)$ with an
uncertainty in luminosity distance due to possible peculiar motion $\delta
v = 400$ km/s of the host galaxy
with respect to the Hubble flow.  Neglecting correlations between errors
(since they are not reported), we
have:
\begin{equation}
\delta H_0 = H_0 \sqrt{ \left ( {\ln 10 \over 5 } \right )^2 \left [ \delta
B^2 + ( b \delta \Delta m_{15} )^2
+ ( R \delta (B-V))^2 \right ]
+ \left [ \left ( {1 \over z} + { {1- q_0} \over 2 } \right ) \delta z
\right ] ^2 }
\end{equation}
\noindent
A weighted average of the 29 values of $H_0 \pm \delta H_0$ then gives a
least-squares value for $H_0$
along with its uncertainty.  The $\chi^2$ for this fit of $H_0$ is added to
the $\chi^2$ for the Cepheid-
calibrated SNe fit of $<M_B^0>$,
both with $b$ and $R$ as parameters.  These are then varied to minimize the
overall $\chi^2$.

When the parameters to be varied enter into the evaluation of the
uncertainty $\delta H_0$ and thus into
the $\chi^2$
of the fit, as they do here, there is a bias favoring larger values of $b$
and $R$ leading to a larger
$\delta H_0$ and thus to a lower $\chi^2$.  We try to eliminate this by
temporarily fixing $b$ and $R$ for
the evaluation of $\delta H_0$ in equation (4) at the first solution
values, then reminimizing and again
fixing them at the new values.  After a few such iterations, stable values
of $b$ and $R$ are found which
are also $\chi ^2$ minima.
In the end, this results in a small decrease in $b$, a substantial decrease
in $R$ (by about 0.6 unit), and a
consequent increase in $H_0$ by about 1.3 units.

\section{RESULTS}

In Table 1, we list the seven SNe from Cepheid-calibrated galaxies and the
three from Cepheid-calibrated
galaxy groups, along with the measured values used in our fits; these are
the distance modulus $\mu$, the
measured apparent magnitude $B$ and the resulting  absolute magnitude $M_B
= B - \mu$ appearing in
equation (3), the decline in $B$ magnitude,
$\Delta m_{15}$, during the first 15 days after maximum, and the color,
taken to be
$B_{\rm max} - V_{\rm max}$.
Where appropriate, the distance modulus incorporates the HST long-exposure
correction of 0.05 as well as
an estimate of the effect of host-galaxy absorption on the Cepheid distance
determination.  Conforming to
the procedure used by Hamuy et al. (1996) for their more distant SNe, both
$B$ and $B-V$ are corrected
for Galactic absorption using the estimates of Burstein \& Heiles (1984),
but no correction is made for
absorption within the host galaxy.  Our method accomodates this host-galaxy
absorption as well as any
intrinsic reddening with the parameter $R$ used in the fits.  Thus the
generally uncertain dust absorption is
effectively corrected for by this procedure. For the 29 Cal\'an-Tololo SNe,
we use values of red-shift, $B$,
$B - V$, and $\Delta m_{15}$ given in Table 1 of Hamuy et al.(1996).

Table 2 presents  the results of our joint fitting of the
Cepheid-calibrated (CC) SNe + Cal\'an-Tololo (CT)
SNe for two data selections.  Shown are the number of fitted supernovae in
each category, the best-fit values
of $H_0$, $<M_B^0>$,
$b$, and $R$, and the individual confidence levels for the best joint fit
of the two subsets of data.  In the first
row we limit the sample to the six directly calibrated CC SNe that also
satisfy the color selection $(B - V) <
0.2 $ (Vaughan et al., 1998) and to the 26 color-selected CT SNe.  In the
second row, the full nearby
sample of 10 CC SNe are jointly fitted with all 29 of the CT SNe.
Both yield, fortuitously, identical values for $H_0 = 62.8$.

It can be seen from Table 2 that the CT data alone fit extremely well in
both cases even though the $b$ and
$R$ parameters are optimized to fit the combined (CT and CC) data.  The
confidence levels for the CT data
are in all cases considerably higher than the most likely value of 0.5.
This reflects the presumed overestimate
of the errors in the CT data set described in Tripp (1998) where these data
alone lead for the 29 (26) SNe to
an unrealistically high confidence level of 0.98 (0.97) for
$b = 0.52\ (0.53)$ and $R = 2.09\  (2.44)$.

In both cases the joint confidence levels are acceptable due in large
measure to these overly good fit of the
CT data.  For the case of six CC SNe, good fits are also realized for the
CC subsample.  But this is not the
case for the 10 CC SNe where their confidence level falls below 1\%.  This
is due to a conflict between the
two most recent CC SNe: 1991T and 1998bu.  They are both somewhat reddened
slow decliners, with the
first being superluminous and the second being subluminous according to
this analysis, so that no $b$ and
$R$ corrections can adequately reconcile the two and at the
same time fit the CT data.\footnote{However, if we fit just the
10 CC SNe alone, then a satisfactory confidence level CL = 0.15 is found,
with $b = 1.56$
and $R = 2.80$.  Alternatively, if one or the other of the two conflicting
SNe is eliminated, then (CC+CT)
fits can be found that are also good CC fits.  Thus, making the color cut,
thereby discarding 98bu and
retaining 91T, yields
$H_0 = 60.8$ with a CC confidence level of 0.36, while selecting just the
directly calibrated SNe and
making no color cut discards 91T and retains 98bu, yielding $H_0 = 64.9$
with a CC confidence level of
0.30.
It is to be expected from the extreme nature of these two SNe, lying, as
they do, far out on either side of the $\chi^2$ distribution, that when one or
the other is eliminated it will substantially impact $H_0$.  It has been
suggested (Fisher et al. 1999) that SN 1991T forms a separate class of
relatively rare superluminous objects not seen among the CT sample. It is
best explained as the result of a merger of two white dwarfs leading to a
super-Chandrasekhar explosion.  SN 1998bu  is a normal supernova, but with
considerable dust obscuration.  Making the conventional
color cut $(B - V) < 0.2$ eliminates it.
Since there is  no evidence coming from interstellar
absorption lines for strong dust obscuration among the 29 CT SNe, whereas
all three of the 10 CC SNe falling outside this cut show clear evidence for
this, in order to minimize bias between the samples the safest procedure is
to apply the color cut to both groups as we do in the CC6/CT26 fit.}

We list in Table 3 the values of $M_B^0$ for each of the Cepheid-calibrated
SNe along with their
residuals, i.e.,
$\chi  = [M_B^0- <M_B^0>] \slash \delta M_B^0 $, for both fits of Table 2.
In Figure 1, we show the
corrected values  (filled circles) and the uncorrected values $M_B$ (open
circles) for the full sample
(CC10/CT29).  These are displayed as a function of $\Delta m_{15}$ in
Figure 1a and as a function of B-V
in Figure 1b.  The strong dependences of the uncorrected open circles on
$\Delta m_{15}$ and $B-V$ are
mostly removed by the best-fit parameters $b = 0.55$ and $R = 2.40$, as
seen in the corrected full circles.
However, as noted above, this fit to a common value of $<M_B^0> = -\!19.44$
has an unsatisfactory
$\chi^2$ with CL = 0.009 due to the conflicting demands of SN 1991T and SN
1998bu.

\section{DISCUSSION}

>From the confidence levels in Table 2, it is evident that the combined
>data from the Cal\'an-Tololo and the
Cepheid-calibrated supernovae can be simultaneously fit in a satisfactory
manner.  However,  the good
confidence level, primarily the result of exceedingly good fits of the CT
SNe by themselves, disguises a
significant difference in the data sets.  This is apparent in Figure 2,
which shows plots of $B - V$ vs. $\Delta
m_{15}$ for the 29 CT SNe (Figure 2a) and the 10 CC SNe (filled circles in
Figure 2b).  The striking
difference between the two groups is the complete absence of unreddened $(B
- V < 0.2)$ events beyond
$\Delta m_{15} = 1.2$ among the 10 CC SNe and their abundance among the 29
CT SNe.  Since $B - V$
and  $\Delta m_{15}$ are both independent of distance, we may include in
Figure 2b six additional nearby
SNe in the Virgo and Fornax clusters (open circles) whose distances are
still uncertain.  This inclusion makes
the two samples more compatible, suggesting that the void in the nearby
sample of 10 may be,
in part, due to
a statistical fluctuation.  Thus, while the 29 CT SNe in Figure 2a display
very little correlation between $B -
V$ and  $\Delta m_{15}$, within the limited statistics of the 10 CC SNe
there is a strong correlation, but
one which is much reduced by the inclusion of the six distance-uncalibrated
SNe.

A possible bias affecting any measurement of the Hubble constant, both in
our two parameter fit and previous one and zero parameter fits, may arise
because the nearby and distant samples come from  galaxies of somewhat
different morphologies.  Since a galaxy must contain a population of young
stars in order to produce Cepheid variables, the CC SNe are generally found
in  spiral galaxies.  (An exception to this is NGC5253, the parent of SNe
1972E and 1895B, which is often classified as an E/SO
peculiar galaxy.  A putative
explanation for its star forming regions is that they are the result of a
galactic merger.).  Thus, apart from SN 1972E, our CC sample comes from
spiral galaxies, whereas the CT sample is about equally divided between
spirals and E, E/SO, and SO galaxies.
Eliminating from CT26 all but the 13 spirals gives for CC6/CT13,
after reminimizing $\chi^2$ for this smaller sample,
a value of $61.9 \pm 2.0$ for $H_0$ compared to $62.8 \pm1.6$
for the full sample CT6/CT26.
Since the former is presumably freer of potential bias we take $H_0 = 62$
as our
best estimate of the Hubble constant
and the associated value of -19.46 for $<M^0_B>$.

We now discuss the uncertainty in $H_0$ arising from a variety of sources.
The 1-$\sigma$ statistical error
in
$\delta H_0$ is found to be $\pm 1.0$.  In addition, there is uncertainty
in $b$ and $R$ which we display in
Figure 3 where $H_0$, as a function of $b$ and $R$, is superposed on the
1-, 2-, 3-, and 4-$\sigma$
contours of the CC6/CT26 fit.
>From this we obtain the 1-$\sigma$ error from the $b$ and $R$ uncertainty
>of $\delta H_0 = \pm1.2$.
Throughout this analysis we have fixed the insensitive deceleration
parameter at $q_0 = -0.45$, found by
making a three parameter
$(b,\  R,\  q_0)$ fit of the 26 CT SNe jointly with the 9 cosmological SNe
of Riess et al. (1998) for which
$B-V$ colors were available.  If we assign an uncertainty of $\pm  0.5$ to
this value of $q_0$, this leads to
an uncertainty
$\delta H_0 =  \mp 0.6$.  Combining these three sources in quadrature
yields $\delta H_0 = \pm 1.7$.

To these uncertainties arising from the SNe Ia analysis must be added a
larger uncertainty coming from the
calibration of the Cepheid variables and their possible metallicity
dependence.  The Madore and Freedman
(1991) distance scale in current use fixes the distance modulus to the
Large Magellanic Cloud (LMC) to be
18.50 mag (50.1 kpc) and scales more distant Cepheid-based galaxy distances
to this value.  Because of the
importance of this number for the determination of the Hubble constant,
much attention has been devoted to
methods for obtaining a more accurate value.  A recent review from a
post-Hipparcos perspective (Walker
1998) recommends a mean modulus of $18.55 \pm 0.10$, found using a variety
of distance indicators.
However, even more recent analyses of an eclipsing binary in the LMC
(Guinan et al. 1998), found during
the OGLE microlensing search, yield a value as low as $18.22 \pm 0.13$ mag
(Udalski et al. 1998).  For this
and other reasons evident in the Walker(1998) summary, we use without
change the distance moduli of
Cepheid-calibrated SNe Ia found by the various observers which are all
based on the 18.50 mag LMC value.
We assign to this value a 1-$\sigma$ error of
$\pm  0.15$ mag, with the usual statistical view that a true distance
modulus  to the LMC differing by twice
that value would be surprising and one differing by 3-$\sigma$ would be
very unlikely.
Using $\delta H_0 = 29(H_0 \slash 62) \delta M_B$ obtained from equation
(1), this uncertainty leads
to a 1-$\sigma$ error of $\delta H_0 =\pm 4.34$.

For some years there has been concern that metallicity differences between
the LMC and the Cepheid-
calibrated galaxies can alter the derived distances to these galaxies.  As
the most easily measured proxy for
metallicity, Kochanek (1997) has collected the logarithmic abundance ratio
$[O/H]$ for a number of the
Cepheid-calibrated galaxies relative to that of the LMC.  For galaxies
hosting SNe Ia, these cover a range of
$[O/H]$ from -0.35 to +0.69.  We have used these to alter the measured
distance moduli $\mu$ by means of
the expression
$\mu^\prime = \mu + \gamma [O/H]$, where the parameter $\gamma$
is varied to obtain a best fit .  For CC6/CT26, the data set for which five
of the Cepheid-calibrated galaxies
have measured values of $[O/H]$, we find $\gamma= +0.23 \pm 0.77$, in
agreement with other recent
findings which fall between
0.14 and 0.31 (Kochanek (1997), Kennicutt et al. (1998), Nevalainen and
Roos (1998)), but showing that
this data set has little sensitivity to $\gamma$.  If we fix $\gamma = 0.3$
and follow the previous procedure
of varying $b$ and $R$
to find a $\chi^2$ minimum, then $H_0$ increases by 0.8 km s$^{-1}$
Mpc$^{-1}$.  Since the question of
a metallicity dependence is still in dispute (see Beaulieu et al. 1997,
Saha et al. 1997, and Saio and Gautschy
1998 for recent differing views), we retain the value of $H_0$ without
correction but assign an uncertainty
in $\gamma$ of  $ {+0.3 \atop -0}$
resulting in $\delta H_0 = {+0.8 \atop -0}$ due to this source.
Combining all these errors in quadrature yields $\delta H_0 = \pm  4.7$ for
the overall uncertainty.

Despite significant differences with other analyses, our value of $H_0
\approx 62$ falls squarely in the
middle of the range spanned by recent determinations using SNe Ia.  These
values of $H_0$, all using the
distance modulus of 18.50 to the LMC, range between 55 and 69 km s$^{-1}$
Mpc$^{-1}$.  One
difference between our analysis and others is that they all seem to use the
nearby approximation to equation
(1), which is tantamount to setting $q_0 = 1$.  If instead they were to use
equation (1) with a negative
$q_0$, as is now apparently required by the cosmological data, then their
values would each increase by
about 1.7 units in cases when the CT SNe are used for the Hubble flow
sample.  The other major difference
is that only our analysis uses two independent parameters ($b$ and $R$) to
standardize the luminosity of the
supernovae; this is required in order to obtain a fit to the CT data with
an acceptable $\chi^2$ (Tripp 1998).
Among the recent analyses using a collection of nearby calibrating
supernovae, Saha et al. (1997), employing
seven Cepheid-calibrated SNe (1895B, 1937C, 1960F, 1972E, 1981B, 1989B,
1990N) and no standardizing
parameters, find $H_0 = 58$.  Suntzeff et al. (1998), using five CC SNe
(1937C, 1972E, 1981B, 1990N,
1998bu) along with the Hubble flow CT SNe and with a one parameter
correction for
$\Delta m_{15}$, obtain $H_0 = 64$.  The difference in $H_0$ found in these
two analyses lies primarily in
the introduction of the $b$ parameter.  Our analysis, involving both $b$
and $R$, reduces $b$ and leads to
an intermediate result.  This is immediately apparent from Figure 3.  As
can be discerned from the figure,
the Saha et al.(1997) $b = R = 0$ fit should yield about 60 for $H_0$.
This is equivalent to their 58 after
correcting for the above effect of $q_0$. Likewise the Suntzeff et al.
(1998) fit with only $b$ as a free
parameter should, after the $q_0$ correction, yield about 66, as can be
seen from the figure for $R = 0$.
Apparently, different selections of Cepheid-calibrated SNe have only a
minor effect on $H_0$.  Thus future
augmentations to the present small number of Cepheid-calibrated SNe will
probably have little impact on
$H_0$.  Assuming that there are no significant observational selection
differences between the CT and CC
samples, only a revision of the Cepheid distance scale will be capable of
altering $H_0$ by more than a few
units.

This work was supported in part by the Director, Office of Energy Research,
Office of High Energy and
Nuclear Physics, Division of High Energy Physics of the U.S. Department of
Energy under contract AC03-
76SF00098 and (for D.B.) an NSF grant AST 9417102.

\clearpage

\centerline{\bf REFERENCES}
\begin{description}
\item Beaulieu J. P., Sasselov D.D., Renault C. et al. 1997, A\&A, 318, L47
\item Branch D., Romanishin W., Baron E. 1996, ApJ, 465, 73
\item Branch D. 1998, ARA\&A, 36, 17
\item Burstein D. \& Heiles, C. 1984, ApJS, 54, 33
\item Fisher, A., Branch, D., Hatano, K. \& Baron, E. astro-ph/980732, MNRS
(in press).
\item Guinan, E. F., Fitzpatrick, L.,E., DeWarf, F.,P. et al. 1999, ApJ,
509, L21
\item Hamuy, M., Phillips, M. M., Schommer, R. A. et al. 1996, AJ, 112, 2398
\item Kennicutt, R. C., Stetson, P. B., Saha, A. et al. 1998, ApJ, 498, 181
\item Kochanek, C. S. 1997, ApJ, 491, 13
\item Lira, P., Suntzeff, N. B., Phillips, M. M. et al. 1998, AJ, 115, 234
\item Madore, B. F., Freedman, W. L. 1991, PASP, 103, 933
\item Nevalainen, J. \& Roos, M. 1998, A\&A, 339, 7
\item Perlmutter, S. et al. 1998, ApJ, in press
\item Phillips, M. M. 1993, ApJ, 413, L105
\item Phillips, M. M., Phillips, A. C., Heathcote, S. R. et al. 1987, PASP,
99, 592
\item Riess, A. G., Filippenko, A. V., Challis, P. et al. 1998, AJ, 116, 1009
\item Saha, A., Sandage, A., Labhardt, L. et al. 1996, ApJ, 466, 55
\item Saha, A., Sandage, A., Labhardt, L. et al. 1997, ApJ, 486, 1
\item Saio, H. \& Gautschy, A. 1998, ApJ, 498, 360
\item Schaefer, B. E. 1996, ApJ, 460, L19
\item Schaefer, B. E. 1998, ApJ, 509, 80
\item Suntzeff, N. B., Phillips, M.M., Covarrubias, R. et al. astro-ph/9811205
\item Szomoru, A. \& Guhathakurta, P., astro-ph/9901422
\item Tanvir, N. R., Shanks T., Ferguson, H. C., Robinson, D. R. T. 1995,
Nat, 377, 27
\item Turner, A., Ferrarese, L., Saha, A. et al. 1998, ApJ, 505, 207
\item Tripp, R. 1998, A\&A, 331, 815
\item Vaughan, T. E., Branch, D., Miller, D. L. \& Perlmutter, S. 1995,
ApJ, 439, 558
\item Udalski, A., Pietrzynski, G., Wozniak, P. et al. 1999, ApJ, 509, L25
\item Walker A., astro-ph/9808336, 1998, in Post Hipparcos Candles,
eds. F. Caputo \& A. Heck (Dordrecht, Kluwer Academic Publ.)
\end{description}

\clearpage

\figcaption{The corrected values $M_B^0$ (filled circles)
for the CC10/CT29 fit and the uncorrected values
$M_B$ (open circles) (top) as a function of
$\Delta m_{15}$ and (bottom) these quantities as a function of
$B_{\rm max}-V_{\rm max}$.
The best-fit corrected value of -19.44 is shown by the horizontal
lines.\label{Fig.1}}

\figcaption{Plot of the distance-independent
quantities $B_{\rm max} - V_{\rm max}$
color vs. $\Delta m_{15}$ (top) for the 29 SNe of the Cal\'an-Tololo
distant sample and (bottom) for the 10 Cepheid-
calibrated SNe (filled circles) and six other nearby SNe (open circles)
from the Virgo and Fornax clusters.  The latter tend to
populate the void of unreddened SNe beyond
$\Delta m_{15} = 1.2$, thereby making the two distributions more
compatible.\label{Fig.2}}

\figcaption{1-, 2-, 3-, and 4-$\sigma$ contours
(corresponding to an increase in $\chi^2$ of 1, 4, 9, 16)
for the CC6/CT26 data set as a function of $b$ and $R$.
Also shown are $H_0 = 60$ and 65 curves as a function of $b$ and $R$.  From
these curves it is apparent why
the $b = 0$, $R = 0$ fit of Saha et al. (1997) yields a lower value than
ours and the $R = 0$ fit of Suntzeff et al. (1998) yields
a larger value.  The $\sigma$ contours show only a weak correlation between
$b$ and $R$ for this data set,
as would be expected,
following the imposition of the color cut from Figure 2.
The $1-\sigma$ uncertainties for $b$ and $R$ are $\pm 0.13$ and
$\pm 0.25$ respectively.\label{Fig.3}}

\clearpage

\begin{deluxetable}{cllllll}
\footnotesize
\tablecaption{Cepheid-Calibrated Supernovae\tablenotemark{a} \label{tbl-1}}
\tablewidth{0pt}
\tablehead{
\colhead{SN} & \colhead{Galaxy}   & \colhead{$\mu(\delta \mu)$}   &
\colhead{$B(\delta B)$}  & \colhead{$M_B(\delta M_B)$} &
\colhead{$\Delta m_{15}(\delta\Delta m_{15})$}     &
\colhead{$B-V[\delta(B-V)]$}
}
\startdata
1937C&IC4182&28.36(9)      (1)&8.80(9)      (6)&-19.56(13)&0.87(10)
(6)&-0.02(6)  (6)\nl
1960F&N4496A &31.13(10)    (1)&11.60(10)  (1)&-19.53(14)&1.06(8)
(11)&0.09(8)      (1)\nl
1972E&N5253&27.94(8)      (1,2)&8.30(14)    (1)&-19.64(16)&0.87(10)
(6)&-0.05(8)  (6)\nl
1974G&N4414&31.41(23)    (3)&12.48(5)    (7)&-18.93(24)&1.11(6)
(7)&0.18(5)      (7)\nl
1981B&N4536 &31.10(13)    (4)&12.03(3)    (6)&-19.07(13)&1.10(7)
(2)&0.10(4)      (6)\nl
1986G\tablenotemark{*}&N5128 &28.13(46)    (1)&11.96(7)
(8)&-16.17(46)&1.73(7)     (12)&0.83(10)    (8)\nl
1989B\tablenotemark{*}&N3627 &30.28(26)    (1)&12.29(6)
(9)&-17.99(27)&1.31(7)     (12)&0.34(5)      (9)\nl
1990N&N4639 &32.03(22)    (1)&12.71(3)    (10)&-19.32(22)&1.07(5)
(2)&0.05(4)      (10)\nl
1991T\tablenotemark{*}&N4527 &31.11(10)    (1) &11.70(2)
(10)&-19.41(10)&0.94(5)     (10)&0.19(3)      (10)\nl
1998bu&N3368&30.37(16)    (5,2)&12.10(3)    (2)&-18.27(16)&1.01(5)
(2)&0.30(4)   (2)\nl
\enddata

\tablenotetext{a}{A list of the 10 Cepheid-calibrated supernovae along with
the input data used in the fits.  The uncertainty for
each quantity is shown in parentheses, followed by the reference for the
measurement.}

\tablenotetext{*} {means the distance is inferred from association with
other galaxies with the distance uncertainty increased
by the projected distance.}

\tablenotetext{}{References: (1) Saha et al. 1997; (2) Suntzeff et al.
1998; (3) Turner et al. 1998; (4) Saha et al. 1996; (5)
Tanvir et al. 1995; (6) Hamuy et al. 1996; (7) Schaefer 1998; (8) Phillips
et al. 1997; (9) Wells et al. 1994; (10) Lira et al.
1998; (11) Schaefer 1996; (12) Phillips 1993.}
\end{deluxetable}

\clearpage

\begin{deluxetable}{crrrrrrrr}
\footnotesize
\tablecaption{Best-fit Solutions\tablenotemark{a} \label{tbl-2}}
\tablewidth{0pt}
\tablehead{
      SNe&   $H_0$&  $<M_B^0>$&      b&       R&\quad&\multispan{3}{\hfil
Confidence level (CL)\hfil}\nl
&&&&&&CT+CC&CT&CC}
\startdata
CC6/CT26 &62.8&-19.42&0.483&1.883&&0.955&0.933&0.719\nl
CC10/CT29 &62.8&-19.44&0.553&2.397&&0.385&0.962&0.009\nl
\enddata
\tablenotetext{a}{The best-fit solutions for the two data sets discussed in
the text.}
\end{deluxetable}

\clearpage

\begin{deluxetable}{lrrrrr}
\footnotesize
\tablecaption{$M_B^0$ and residuals.\tablenotemark{a} \label{tbl-3}}
\tablewidth{0pt}
\tablehead{
Supernova&\multispan{2}{\hfil CC6/CT26\hfil}&\quad&\multispan{2}{\hfil
CC10/CT29\hfil}\nl
 &$M_B^0\quad$&Residual&&$M_B^0\quad$&Residual
}
\startdata
1937C&-19.44 (18)&-0.10&&-19.41 (20)&0.15\nl
1960F&-19.70 (21)&-1.36&&-19.75 (24)&-1.27\nl
1972E&-19.46 (23)&-0.18&&-19.42 (26)&0.90\nl
1974G&-19.30 (26)&0.47&&-19.39 (27)&0.18\nl
1981B&-19.28 (16)&0.87&&-19.34 (17)&0.63\nl
1986G*&&&&-18.54 (52)&1.73\nl
1989B*&&&&-18.95 (30)&1.67\nl
1990N&-19.39 (24)&0.11&&-19.45 (24)&-0.03\nl
1991T*&&&&-19.80 (13)&-2.83\nl
1998bu&&&&-18.97 (19)&2.49\nl
\enddata
\tablenotetext{a}{Note: Values of $M_B^0$ with its error $\delta M_B^0$ and
the residuals
$\chi  = [M_B^0 -<M_B^0>] \slash \delta M_B^0$ for each Cepheid-calibrated
supernova for the
two fits discussed in the text.
}
\end{deluxetable}

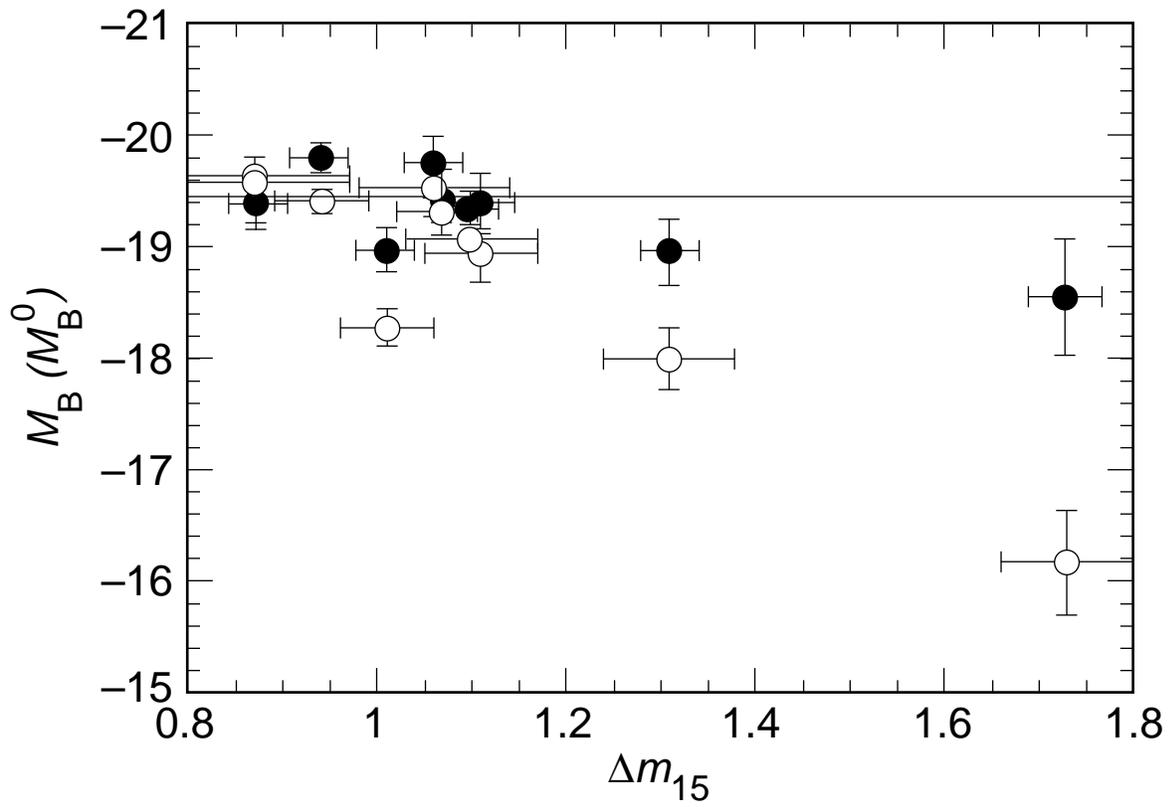
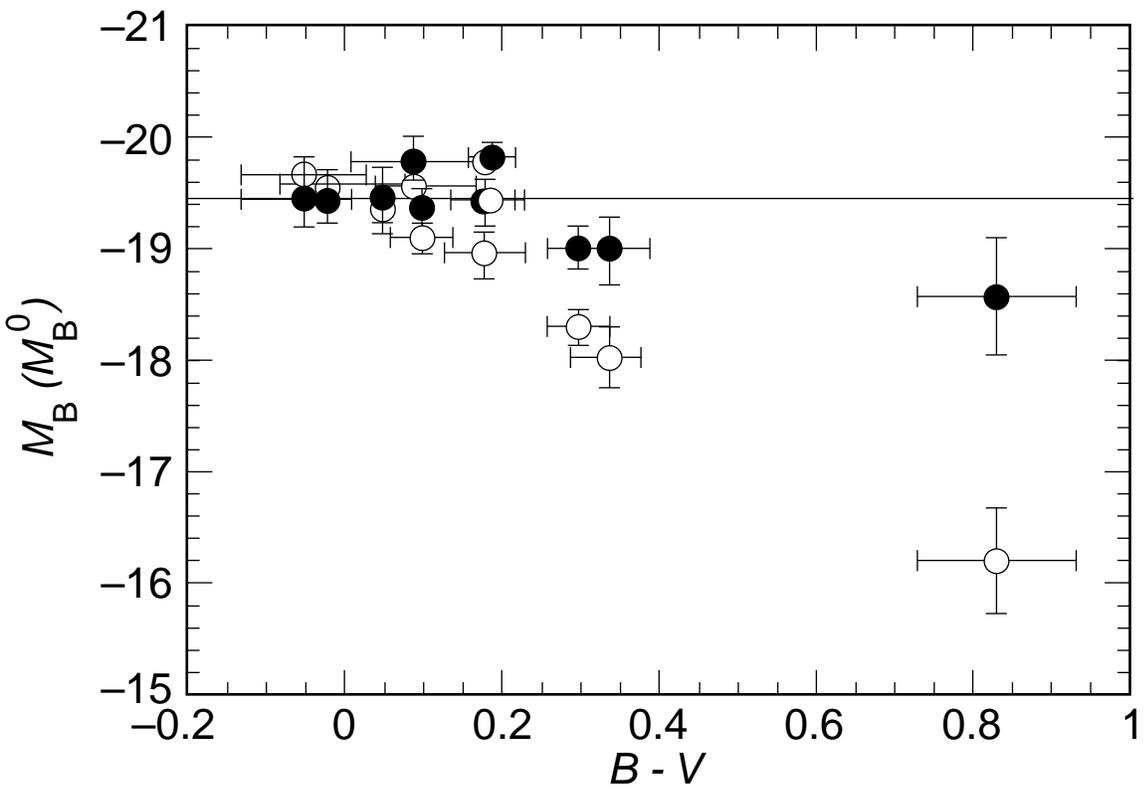
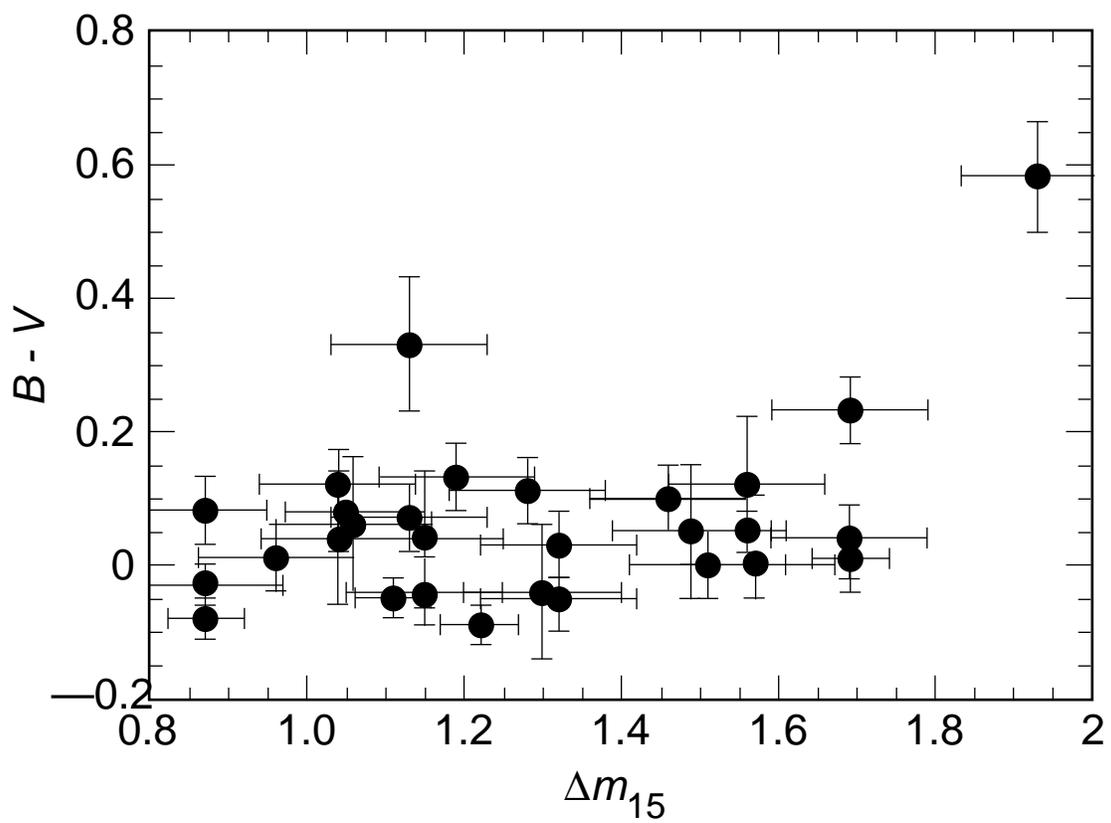
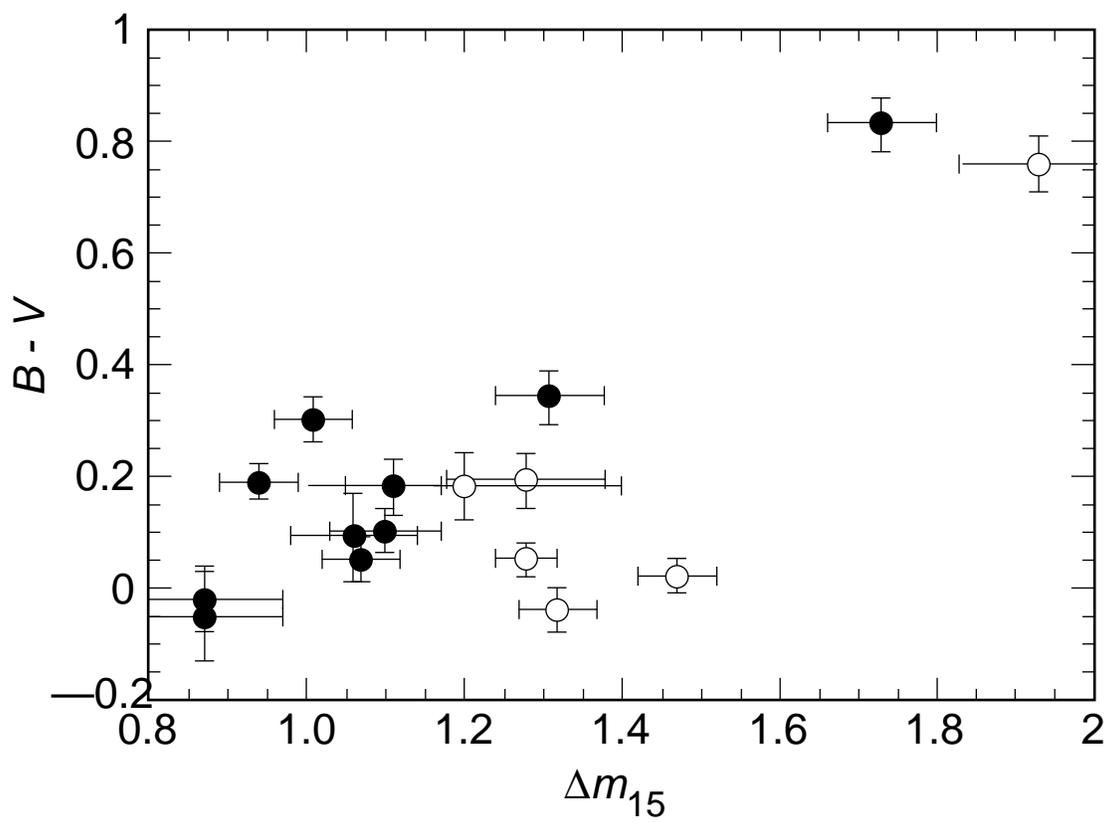
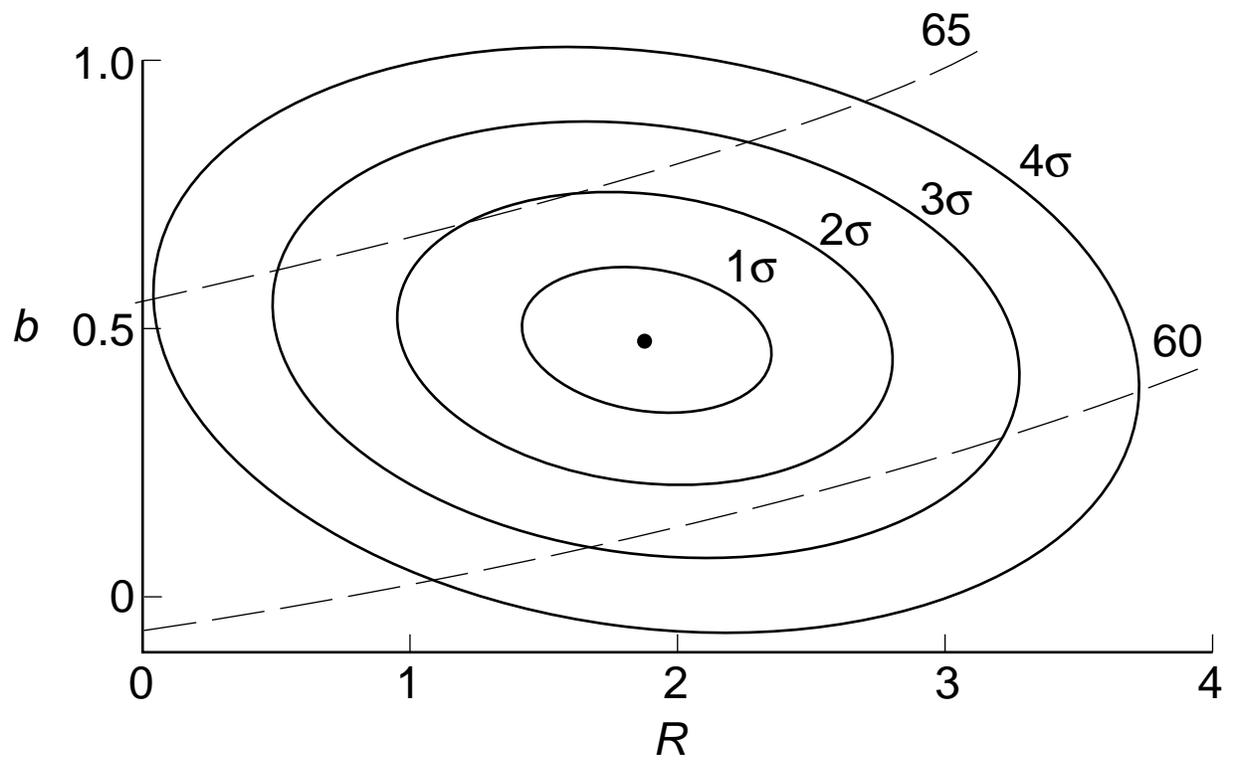
\end{document}